%% file: Manuscript.tex
\documentclass[12pt]{article}  %
 \usepackage{geometry}
 \usepackage{setspace} %
\usepackage[table]{xcolor} %
\usepackage{makecell}

\renewcommand\cellalign{rt}
\renewcommand\arraystretch{0.9} %
\usepackage{amsmath,bm}
\usepackage{algorithm}
\usepackage{algpseudocode}
\usepackage{tikz}
\usepackage{hyperref}
\usepackage{xurl}
\usepackage{booktabs}
\usepackage{tabularx}
\usepackage[svgnames]{xcolor}
\usepackage[flushleft]{threeparttable}
\usepackage{footnote}
\usepackage{makecell}

\usepackage{xcolor}

\usepackage{siunitx} %
\renewcommand{\arraystretch}{0.9} %

\usepackage{listings}
\usepackage{tcolorbox}
\tcbuselibrary{breakable}

\usepackage{natbib}
 \bibpunct[, ]{(}{)}{,}{a}{}{,}%

\usepackage{multibib}
\newcites{appendix}{Appendix References}
\usepackage{cleveref}
\usepackage{appendix}

\title{When Machines Meet Each Other: Network Effects and the Strategic Role of History in Multi-Agent AI Systems}

\author{
	Yu Liu, Wenwen Li, Yifan Dou, Guangnan Ye\\
	Fudan University\\
	yuliu23@m.fudan.edu.cn, \{liwwen, yfdou, yegn\}@fudan.edu.cn
}
\date{} 

\begin{document}

\maketitle

\begin{abstract}
As artificial intelligence (AI) enters the agentic era, large language models (LLMs) are increasingly deployed as autonomous agents that interact with one another rather than operate in isolation. This shift raises a fundamental question: how do machine agents behave in interdependent environments where outcomes depend not only on their own choices but also on the coordinated expectations of peers? To address this question, we study LLM agents in a canonical network-effect game, where economic theory predicts convergence to a fulfilled expectation equilibrium (FEE). We design an experimental framework in which 50 heterogeneous GPT-5–based agents repeatedly interact under systematically varied network-effect strengths, price trajectories, and decision-history lengths. The results reveal that LLM agents systematically diverge from FEE: they underestimate participation at low prices, overestimate at high prices, and sustain persistent dispersion. Crucially, the way history is structured emerges as a design lever. Simple monotonic histories—where past outcomes follow a steady upward or downward trend—help stabilize coordination, whereas non-monotonic histories amplify divergence and path dependence. Regression analyses at the individual level further show that price is the dominant driver of deviation, history moderates this effect, and network effects amplify contextual distortions. Together, these findings advance machine behavior research by providing the first systematic evidence on multi-agent AI systems under network effects and offer guidance for configuring such systems in practice.
\end{abstract}

\vspace{0.5\baselineskip}

\section{Introduction}\label{sec:Intro}

With artificial intelligence entering the agentic era, large language models (LLMs) are no longer confined to serving as isolated tools but are increasingly deployed as autonomous entities that interact, compete, and adapt to one another in complex environments, such as finance, logistics, and digital platforms. For example, algorithmic trading systems now battle in milliseconds to exploit fleeting opportunities; recommender systems interactively shape consumer demand; and generative agents collaborate—or clash—on online platforms to influence collective outcomes \citep{rahwan2019machine, anthis2025llm, dou2025ai}. In all these cases, what matters is not a single model’s performance in isolation, but the emergent dynamics of multi-agent interaction. Understanding this interdependent behavior is rapidly becoming one of the most pressing and theoretically consequential frontiers in economics and information systems.

Studying interdependent LLM agents poses a distinctive theoretical challenge: unlike a standalone predictor, an LLM agent in a networked environment must forecast not just outcomes but \textit{other agents’ forecasts of others’ forecasts}. Classical economics resolves this infinite recursion by invoking equilibrium beliefs to solve compact fixed‐points that discipline expectations into action. In settings with network effects, the fulfilled expectation equilibrium (FEE) instantiates this logic: realized participation matches a common expectation, which delivers a tractable benchmark for coordination under interdependence \citep{katz1985network}. Decades of work show that such equilibrium reasoning can organize human behavior in the presence of social influence, cascades, and increasing returns \citep{bikhchandani1992theory,arthur1989competing,camerer2004cognitive,boudreau2021promoting}. Whether machine agents—whose “beliefs” emerge from sequence prediction and memory rather than explicit fixed‐point computation—can approximate analogous coordination is an open question at the intersection of information systems, economics, and machine behavior \citep{rahwan2019machine,fan2024can,sun2025game}.

Network effects are not merely another strategic externality; they are a many-to-many dependency that magnifies the gulf between economic equilibrium concepts and contemporary LLM architectures. In economic theory, FEE presumes that all agents (i) share a common model of the environment and of others, (ii) process identical information, and (iii) compute the same fixed point, yielding uniform expectations and self-fulfilling outcomes. In practice, large multi-agent systems could violate each presumption: information is path-dependent, attention is finite, and inference is implemented by stochastic token-by-token prediction rather than common-knowledge computation. Recent work confirms the tension: LLMs can mimic rational play in simple dyadic or matrix games \citep{silva2024large,fontana2025nicer}, but they often fail in environments with heterogeneity or dynamic interdependence \citep{deng2025can,huang2024far}. Focusing on network effects therefore offers a uniquely stringent testbed: it requires alignment on a shared expectation about everyone’s participation, at scale. By probing this setting, our study isolates whether—and how—the equilibrium logic that disciplines human interaction in economic theory transfers (or fails to transfer) to LLM-based communities.

To investigate, we construct a canonical network-effect game with economic agents and then reconfigure the environment for LLM agents. In our implementation, 50 autonomous LLM agents are each assigned heterogeneous standalone values and are exposed to external conditions including prices and network effects. Importantly, while classical economic agents make choices based on expected payoffs, it is unknown whether LLM agents would explicitly ``solve'' for equilibrium. Instead, we elicit their \textit{predictions} of how many agents will participate at a given price, and treat these forecasts as their reported ``expectations.'' By varying both the strength of network effects and the structure of historical information, we isolate how network effects and history input shape these forecasts. The design allows us to compare the classical FEE benchmark derived from economic theory and the collective behavior of LLM agents as revealed through simulation.

Our findings reveal that LLM agents behave in ways that depart systematically from textbook economic predictions. In static settings without history, they fail to replicate the fulfilled expectation equilibrium: at low prices, agents consistently underestimate participation, while at high prices they overestimate it. Rather than converging, their forecasts remain dispersed. This divergence becomes more pronounced in environments with stronger network effects. Although the direct effect of network strength is not statistically robust, its presence magnifies deviations generated by contextual factors, leading to wider gaps between LLM forecasts and theoretical equilibria.

Interestingly, when LLM agents can access the history of outcomes from previous rounds, their ``expectations'' shift, but convergence to FEE still does not emerge. A distinctive feature of our design is that history can be organized and presented in different ways, giving us a unique opportunity to explore how memory might alter coordination—something almost impossible to test systematically in human settings. While our exploration is necessarily preliminary, the results are suggestive. Monotonic histories provide stabilizing cues: longer history reduces dispersion and nudges collective forecasts closer to the benchmark. By contrast, non-monotonic histories do not effectively improve the coordination of LLM agents' ``beliefs'', amplifying divergence and reinforcing path dependence. Taken together, these explorations point to a central conclusion: History plays a critical role in shaping LLM agents’ strategic behavior, while they are treated as ``sunk'' in the benchmarking economic theory.

To explore with greater granularity, we turn to individual-level regressions to quantify the forces driving LLM agents’ deviations. The analysis shows that network effects do not act as an independent main force; instead, their influence lies in reshaping and conditioning other drivers of behavior. External incentives such as price and internal heterogeneity in standalone values both emerge as systematic sources of deviation, while history plays a moderating role by dampening sensitivity to extremes and curbing dispersion. Network effects, notably, amplify these contextual pressures, magnifying price-driven distortions and conditioning the impact of heterogeneity. This layered picture sharpens a central theoretical insight: in multi-agent systems, at least with today’s LLMs, equilibrium reasoning does not emerge endogenously. Rather, it is the interplay of external incentives, internal heterogeneity, and historical context—knit together by network effects—that drives LLM agents' ``expectations''.

These results carry intertwined theoretical and practical significance. Theoretically, they show that equilibrium reasoning—a cornerstone of economics—does not arise in a similar fashion in LLM-based systems. Expectations prove contingent, history-dependent, and sensitive to architecture, calling for equilibrium concepts that are history-aware and tailored to machine cognition. Practically, our findings caution that outcomes in AI-driven markets and platforms depend not only on incentives and payoffs, but also on how systems are configured. Elements often treated as secondary in human economics, such as the framing of historical trajectories, emerge as primary determinants of machine behavior. Designing and governing AI collectives therefore requires equal attention to incentives, network interdependencies, and the informational scaffolding through which agents process the past.

At a broader level, our study provides the one of the first systematic, quantitative evidences of machine behavior in many-to-many multi-agent environments shaped by network effects. By comparing to the FEE benchmark with experimental simulations and regression analyses, we uncover the pattern on how LLM agents diverge from classical equilibrium reasoning and how history matters. This opens a new conversation for a research agenda at the interface of economics and AI -- our results suggest that equilibrium models must be re-examined in light of machine cognition, as the interdependence is a non-negligible lens for understanding the future of autonomous decision-making.

The remainder of the paper proceeds as follows. Section~\ref{sec:lit} reviews related literature. Section~\ref{sec:game} introduces the benchmark economic model. Section~\ref{sec:configuration} describes how we configure LLM agents and design the experiments. Section~\ref{sec:results} presents the simulation findings, and Section~\ref{sec:regression} reports regression analyses. Section~\ref{sec:conclusion} concludes. Technical details are provided in the Appendix \footnote{The appendix is not avaliable in this version but can be requested from the authors.}, including robustness checks with Qwen-Plus, while the main text focuses on the GPT-5–based LLM agents.

\section{Related Literature}\label{sec:lit}

Three streams of literature are closely related to our work: LLMs in social sciences, LLMs in game-theoretic contexts, and the classical economic literature on network effects and fulfilled expectations.

\subsection{Large Language Models and Their Applications in Social Science}\label{subsec.lit.llm}

LLMs have rapidly evolved from linguistic tools into intelligent agents that increasingly augment or even substitute for human decision-making across diverse domains. Originally developed for natural language processing tasks such as translation and text generation, LLMs now demonstrate emergent capabilities in reasoning, planning, and strategic interaction \citep{achiam2023gpt}. These advances have attracted growing attention from social scientists, who view LLMs both as \textit{objects of study} that exhibit behavioral patterns akin to human cognition (``social science for AI''), and as \textit{instruments} for conducting large-scale behavioral research (``AI for social science'') \citep{rahwan2019machine, xu2024ai, horton2023large}.  

For example, LLMs have been employed to approximate consumer preferences, trust formation, and social influence \citep{xie2024can, fan2024can}; to serve as artificial participants in experimental economics and psychology \citep{willis2025will}; and to simulate decision processes in organizational settings \citep{chen2023emergence, huang2024far}. Beyond the laboratory, LLM-based agents are increasingly integrated into customer service, marketing, and digital platforms, where their design features shape user experience and collective outcomes. Recent IS work highlights how emotional expression by bots alters customer evaluations \citep{han2023bots}, how disclosing human involvement shifts perceptions of hybrid service agents \citep{gnewuch2024more}, and how affinity and trust in digital humans influence adoption \citep{seymour2025less}. Complementary perspectives emphasize the need for systematic frameworks for embedding AI agents into socio-technical systems \citep{abbasi2024pathways, yoo2024next}.

Collectively, this stream underscores that LLM-based agents are not neutral computational devices but actors whose behavior is contingent on environmental framing, historical cues, and system design. Building on this logic, our study adopts a deliberately minimal configuration: rather than embedding agents in complex socio-technical contexts, we focus on one of the simplest interdependent settings governed by network effects. This stripped-down design provides a clean and transparent environment in which to isolate the mechanisms at stake and to examine how LLM agents behave when strategic interdependence alone drives outcomes.

\subsection{Strategic Behavior and Strategic Reasoning of LLM Agents}\label{subsec.lit.agent}

A second stream focuses on the strategic behavior of LLM agents in game-theoretic settings. Within computer science and AI, LLMs have been evaluated in canonical games such as the Prisoner's Dilemma, Stag Hunt, and matrix coordination games. These studies show that LLMs can replicate basic patterns of cooperation or competition, often adapting to contextual cues embedded in prompts \citep{lore2024strategic, gandhi2023strategic, akata2025playing}. Yet systematic analyses reveal limits: \citet{fan2024can} find that LLMs struggle with recursive reasoning, while \citet{huang2024far} show that decisions in sequential games are heavily shaped by preceding outcomes. \citet{guo2024economics} demonstrate that historical information improves coordination in public goods and ultimatum games, but deviations from equilibrium play persist. Other recent contributions expand to larger populations and more complex simulations. \citet{anthis2025llm} and \citet{taillandier2025integrating} discuss the promise and challenges of LLM-based social simulations, while \citet{karten2025llm} propose an ``LLM economist'' to explore mechanism design in large collectives. 

Our study advances this literature by moving beyond dyadic or small-group games to analyze many-to-many interdependencies with recursive expectations—settings that are theoretically fundamental but empirically underexplored. This gap motivates our focus on LLM agents in network-effect environments, where their ability—or inability—to coordinate expectations offers critical insight into the nature of machine behavior under interdependence.

\subsection{Network Effects, Expectations, and Fulfilled Equilibria}\label{subsec.lit.ne}

Network effects have long been recognized as central to economics, strategy, and information systems. When an agent's payoff increases with the number of peer adopters, positive feedback loops generate multiple equilibria, path dependence, and tipping dynamics \citep{arthur1989competing, bikhchandani1992theory}. To resolve the recursive expectations problem in such environments, classical models employ the \textit{fulfilled expectation equilibrium} (FEE), in which realized adoption equals anticipated adoption \citep{katz1985network}. This concept embodies the rational-coordination principle: each agent anticipates the collective, and the collective in turn confirms that anticipation.  

FEE has since become the dominant analytical device for studying network effects. Foundational works such as \citet{farrell1985standardization} and \citet{katz1986technology} showed how expectations, compatibility, and standardization jointly determine adoption dynamics. \citet{economides1996economics} synthesized the economics of network industries, demonstrating how pricing, compatibility, and equilibrium selection interact. Subsequent researchers have extended FEE to diverse domains, including platform competition, social networks, and information cascades, where fulfilled expectations provide tractable benchmarks for collective outcomes. In management research, \citet{conner1995imitated} demonstrated how imitation strategies under network effects accelerate adoption and strengthen competitive advantage, while \citet{jing2007network} analyzed monopolistic coverage under network externalities.  

Whether FEE remains applicable when extended from human decision-makers to AI agents, however, is far from clear. Classical analyses rely on the assumption that actors are forward-looking and capable of recursive reasoning, aligning individual beliefs with collective outcomes through fixed-point logic. LLM agents, by contrast, generate expectations through statistical associations and in-context memory rather than explicit recursive calculation. This distinction raises an open theoretical question: can fulfilled expectations, so central to human-centered models of network effects, also describe equilibrium reasoning in groups of LLM agents? Our study provides one of the first systematic attempts to answer this question by testing the robustness and limits of classical equilibrium logic in the emerging domain of multi-agent AI systems.  

\section{Benchmark: A Network-Effect Game with Economic Agents}\label{sec:game}

As a benchmark, we begin with a canonical example of a network-effect game in which six economic agents must decide (hypothetically) whether to attend a conference. Each scholar is indexed by $i \in \{1,2,\ldots,6\}$ and is assumed to know the total number of potential participants (i.e., 6), the available action set $\{\text{Attend}, \text{Not Attend}\}$, and her own parameters: a standalone valuation $\theta_i = i$ that captures the standalone value of attending (e.g., visiting the conference venue and sight-seeing), a fixed cost $p$ that reflects travel and registration expenses, and a common coefficient $\beta$ that measures the strength of network effects (e.g., the benefit from interacting with other participants). The payoff function for each scholar $i$ is
\begin{equation*}
U(\theta_i) = \theta_i + \beta N - p, 
\end{equation*}
where $N$ denotes the total number of other attendees. A scholar $i$ will choose to attend if and only if $U(\theta_i) \geq 0$. 

The main challenge in this setting is that $N$ is not observable ex ante, forcing each economic agent to form expectations about the participation decisions of others in order to make her own choice. This circularity in reasoning is resolved in classic economic theory through the concept of the \textit{fulfilled expectation equilibrium} (FEE). In an FEE, all agents coordinate on a common expectation, and the realized number of participants is consistent with that expectation \citep{katz1985network}. This solution concept characterizes equilibrium under network effects and provides a tractable analytical benchmark. 

To illustrate, consider the six-scholar example with $p=4.4$ and $\beta=0.5$. Under FEE, each agent can calculate that the last scholar willing to participate is $i=3$. Specifically, if scholars 3 through 6 attend (i.e., $N=4$), then $U(\theta_2) = 2 + 0.5 \times 4 - 4.4 < 0$, which implies that scholar 2 does not attend. Because every scholar can perform the same calculation, they converge to the identical expectation that only scholars 3--6 will attend, and the outcome indeed fulfills that expectation--mathematically, a fixed point. 

In our study, this classical framework serves as the baseline against the results from the same game with six LLM agents, which will enable us to examine whether machine-driven systems replicate or deviate from the theoretical predictions established in the classic economic literature.

\section{Configuring the Network-Effect Game with LLM Agents}\label{sec:configuration}

Building on the benchmark, we now investigate the case with LLM agents. The central motivation is that classical economic agents are typically forward-looking and treat past outcomes as sunk, whereas LLM agents inherently rely on historical information to form their predictions of the future. This discrepancy implies that LLM agents may deviate from the fulfilled expectation equilibrium. To capture this distinction, we extend the canonical network-effect game into a large-scale experiment with 50 agents. We distinguish between two scenarios. In the \textit{Non-Repeated Game (Static)}, agents face a single-shot decision without access to prior outcomes. In the \textit{Repeated Game (Dynamic)}, games unfold recursively with evolving prices, enabling LLM agents to incorporate past expectations, realized participation, and payoffs into their subsequent decisions. By contrasting these two settings, we isolate how LLM agents exploit historical information.

Each of the 50 independent agents is assigned a unique standalone value, $\theta_i \in \{0,\dots,49\}$, mirroring the heterogeneity in the benchmark game. We consider two levels of network effects, $\beta \in \{0.25, 0.75\}$. To explore the role of decision history, we designed five price sequences under both static and dynamic settings. For each network-effect level, six theoretical equilibrium prices were computed, corresponding to 0, 10, 20, 30, 40, and 50 participants under with economic agents. 

In the \textit{static} scenario, the price is fixed at one of the six theoretical levels, and the experiment is repeated across all points. This design isolates agent behavior at equilibrium benchmarks under weak and strong network effects. In the \textit{dynamic} scenario, each experiment spans six rounds, with price sequences constructed to cover the same theoretical equilibrium points. We test four representative trajectories: (i) \textit{decreasing} sequences, where prices rise steadily (thus the number of participants is supposed to decrease); (ii) \textit{increasing} sequences, where prices fall steadily; (iii) \textit{converging} sequences, where prices begin at an extreme and gradually approach the mean; and (iv) \textit{diverging} sequences, where prices start near the mean and move outward toward extremes. The monotonic sequences mimic sustained market growth or contraction, while the non-monotonic sequences capture more volatile dynamics with overshooting or oscillation.

Overall, the experiment follows a $2 \times ( 1+ 3 \times 4)$ design: two network-effect strengths, static scenario, and a dynamic scenario with three decision-history settings (short, medium, long) crossed with four market dynamics, yielding 26 unique conditions. This systematic setup allows us to disentangle how network effects, historical dependence, and market trajectories interact in shaping agents’ collective behavior and aggregate outcomes. The calculations of equilibrium prices and sequence construction appear in Appendix.

\subsection{Experimental Workflow Design}\label{subsec:workflow}
\input{contents/figures/finite_state_machine}

To justify our implementation choices, we formalize the experiment as a finite state machine (FSM), which is a routine way to specify multi-agent, discrete-event processes for transparency and replication \citep{zuzak2011finite, stodden2016enhancing, rosales2021composable}. This abstraction elevates analysis from individual agents to the full system (agents interacting with an environment), makes timing and information flow explicit, and turns the protocol into an auditable artifact that others can reproduce.

To align with economic modeling, our workflow mirrors the standard timing of a simultaneous-move game with public signals and privately known types. In the static case, agents best-respond once to current prices and public information, treating past outcomes as sunk; in the dynamic case, repeated play exposes history-dependence in expectation formation—precisely where LLM agents may diverge from forward-looking rationality \citep{katz1985network}. This design ensures that any deviations from the fulfilled expectation equilibrium arise from agents’ use of history rather than from an idiosyncratic protocol.

Formally, we encode the protocol as a discrete-event model $\mathcal{M}=(S,E,\delta,S_0)$, where states $S$ capture mutually exclusive phases, events $E$ are instantaneous triggers, the transition rule $\delta:S\times E\to S$ advances the workflow, and the unique initial state $S_0$ guarantees identical initialization across runs. Figure~\ref{fig:fsm} depicts this logic and provides a compact specification for implementation and audit.

Operationally, five states cover one round of play while cleanly separating environment control from agent logic. In $S_0$ (Initialization), the environment fixes global parameters (e.g., $\beta$, the price sequence, and the utility definition) and assigns each agent its private value $\theta_i$. In $S_1$ (Information Broadcast), the environment publishes the current price $p_t$ and the previous round’s public feedback (e.g., $N_{t-1}$). In $S_2$ (Agent Decision-Making), agents move simultaneously and independently from expectations to a binary action given their private type and public signals. We formalize the decision-making process into a sequence of three distinct sub-processes: Information Gathering (including $\beta$, the price sequence, and the utility definition, $N_{t-1}$), Expectation Formation \& Utility Calculation and Decision Execution. In $S_3$ (Outcome Aggregation \& Payoff), the environment aggregates actions to obtain $N_t$ and computes payoffs from the utility function. Finally, $S_4$ (Termination) absorbs the process once all conditions are satisfied, closing the run deterministically.

Events induce a fixed progression that clarifies timing and information. The environment triggers \emph{Experiment Start} ($S_0\!\rightarrow\!S_1$) and \emph{Broadcast Complete} ($S_1\!\rightarrow\!S_2$); agents jointly trigger \emph{All Decisions Complete} ($S_2\!\rightarrow\!S_3$); the environment triggers \emph{Results Calculated} within $S_3$ and, when the price sequence is exhausted, \emph{Termination Met} ($S_3\!\rightarrow\!S_4$). The workflow thus cycles $S_1\!\rightarrow\!S_2\!\rightarrow\!S_3$ over prices in the dynamic setting, or exits to $S_4$ after a single price in the static setting.

This FSM formalization serves three aims central to Information Systems (IS) research practice. First, it preserves the economic game’s timing and information structure, maintaining internal validity relative to the benchmark. Second, it makes history-dependence an explicit, manipulable treatment, which is essential for studying learning and expectation formation by LLM agents. Third, it enhances transparency and replicability by constraining non-determinism and documenting every state transition \citep{stodden2016enhancing}. Together, these features provide a standard, defensible scaffold for verification, robustness checks, and future extensions.

\subsection{Experiment Setup}

We employed state-of-the-art LLMs as the foundation for the agents in our simulation. The overall environment was engineered with a strong emphasis on reproducibility and rigorous evaluation, ensuring that our results are transparent, verifiable, and not dependent on ad hoc implementation choices.

The choice of models reflects the need to capture variation across both commercial and open-source architectures. We used three widely adopted LLMs as the backbone for our agents: OpenAI’s GPT-5 and Tongyi Qianwen’s Qwen3-Plus. GPT-5 represents the current peak of commercial performance and provides a natural benchmark for advanced reasoning. Qwen3-Plus offers a balanced and reliable commercial alternative. Taken together, these models allow us to test whether observed behaviors are tied to specific model families or generalize across architectures.

The interaction between agents and the environment followed a structured communication protocol. All exchanges were based on a predefined JSON schema to guarantee consistency and machine readability. The environment acted as a central coordinator: broadcasting the current price and past outcomes, collecting agent responses, and aggregating results. Each agent’s identity and private parameters were encoded in its system prompt. At every round, the environment translated structured information (e.g., price $p_t$, last period’s participation $N_{t-1}$) into a text-based prompt that the LLM could process. Agents returned their choices in a JSON-compliant format, which ensured that outputs could be parsed and analyzed reliably. This design mirrors the way economic experiments communicate payoff-relevant signals and collect decisions, while also adhering to best practices in multi-agent computational experiments.

Finally, the decoding configuration was standardized to balance predictability with exploratory behavior. We set the temperature parameter to 0.7 for all models, a common practice in computational social science that encourages both precise reasoning and limited variability \citep{chen2023emergence,horton2023large}. All other decoding parameters (e.g., top-p, frequency penalty) were held constant across models to maintain comparability. Configuration details are provided in the Appendix, including decoding strategy, parameter settings, and prompt engineering for dynamic agent-system interaction. To further test robustness, we conducted supplementary trials with a lower temperature of 0.35. The results are presented in Appendix. The consistency of the results under this adjustment suggests that our conclusions are not sensitive to specific decoding configurations, thereby strengthening the reliability of our experimental findings.

Overall, this setup ensures that LLM agents face the same informational environment as economic agents in the benchmark game, with the key distinction being how expectations are formed. Any deviation from theoretical predictions can thus be attributed to the way LLMs process and use historical information, rather than to artifacts of model selection, communication design, or decoding parameters.

\section{Experimental Results and Analysis}\label{sec:results}

We first analyze the \emph{static} scenario in Section \ref{subsec.static}, where LLM agents have no access to history and each time must decide using only the information at hand. We then turn to the \emph{dynamic} scenario in Section \ref{subsec.dynamic}, where LLM agents can condition on history and we systematically vary the informational environment via four price trajectories (monotone increasing/decreasing and non-monotone converging/diverging). For each scenario and trajectory, we present figures that report the distribution of agents’ expectations alongside the theoretical benchmark, highlighting where LLM agents' behavior aligns with or departs from economic theory. After the visualization evidence, we introduce quantitative measures—most notably aggregate fit metrics (RMSE)—to summarize deviations from the benchmark and to compare performance across network-effect strengths, models, and history configurations.

\subsection{Static Scenario Results}\label{subsec.static}

In the static case, LLM agents are given no historical data. We examined two state-of-art LLM types (GPT-5 and Qwen3-Plus). The findings are presented in Figure~\ref{fig:results_benchmark}, with each column corresponding to a different LLM and each row representing a different network effect strength ($\beta=0.25$ or $\beta=0.75$). The blue box plots show the expected number of participants for a group of agents under each price ($p$), with the compactness of the box reflecting decision consistency across agents (i.e., smaller boxes indicate more similar expectations). The dark blue dashed line connects the means of these box plots, tracing the trend of the group’s average expectation. The red dashed line depicts the theoretical equilibrium solution under the fulfilled expectation equilibrium (FEE), which serves as the benchmark.

\input{contents/figures/results_benchmark}

Figure~\ref{fig:results_benchmark} illustrates the experimental results in detail. Along the horizontal axis, higher prices correspond to fewer expected participants under FEE, as reflected by the downward-sloping red dashed line. We selected five representative price points ($\{12.49, 19.99, 27.49, 34.99, 42.99, 49.99\}$), which map to equilibrium participation levels of $\{50, 40, 30, 20, 10, 0\}$ under FEE.\footnote{Decimal values such as $.99$ are used instead of integers to avoid ambiguity in cases where $U(\theta)=0$, in which an agent is indifferent between participating or not.} Because all economic agents under FEE converge on the same expectation, each price point corresponds to a single value along the red dashed line.

By contrast, the LLM agents exhibit heterogeneity in their expectations. At each price, the spread of responses forms a box (covering the interquartile range), with the blue dashed line marking their mean. Relative to the theoretical benchmark, LLM agents are systematically pessimistic at low prices (their mean expectation lies below the red line) and systematically optimistic at high prices (their mean lies above the red line). This pattern is consistent across both models.

Finally, when the strength of network effects increases from $\beta=0.25$ (top row) to $\beta=0.75$ (bottom row), the deviation from the benchmark becomes larger. This suggests that stronger network effects exacerbate the divergence between LLM behavior and theoretical predictions, a trend we quantify more rigorously later.

\subsection{Dynamic Scenario Results}\label{subsec.dynamic}

We now turn to the dynamic setting, where LLM agents are given access to historical information when making their decisions. This design allows us to test whether and how LLM agents leverage the history. To operationalize, we systematically vary both the trajectory of prices and the length of the history window available to the agents. Specifically, four types of price trajectories are considered: monotonic sequences in which prices either increase steadily from low to high or decrease from high to low, and nonmonotonic sequences in which prices either converge toward a mean or diverge toward extremes. In each case, we ask four central questions: (i) does more history lead to more similar expectations among agents (i.e., tighter boxes), (ii) does it bring the mean expectation closer to the FEE benchmark, (iii) are there systematic biases depending on the direction of the trajectory, and (iv) does a stronger network effect make convergence to FEE more difficult? The analysis focuses on GPT-5,  with robustness checks for Qwen3-Plus reported in Appendix.

An important feature in the design above is how the history window is configured. In each figure, moving from the left panel to the right, the length of the memory available to agents increases: With a short window, agents can only recall the most recent round (e.g., the immediately smaller price just observed, if any). With a medium window, they can recall the last three rounds (if any), while with a long window they can perfectly remember the entire history of the game up to that point. Without loss of generality, we set these three levels as short = 1, medium = 3, and long = 6 rounds. Besides, each row of panels represents a different degree of network effects, with the top row under $\beta=0.25$ and the bottom row under $\beta=0.75$.

\subsubsection{Monotonic Sequences}

We begin with the increasing-price trajectory (see Figure \ref{fig:results_gpt_5_increasing}). In this setting, the price given to the LLM agents rises steadily. From left to right, each panel shows how extending the history window changes expectations. With only a short memory (one round, see Figure \ref{fig:results_gpt_5_increasing}-A and \ref{fig:results_gpt_5_increasing}-D), the dispersion of expectations is large, and the mean deviates visibly from the FEE benchmark. As the history window increases to three rounds, the dispersion narrows substantially, and the mean moves closer to the FEE line. Extending the history window further to six rounds continues this trend, though the marginal improvement is small, suggesting diminishing returns from providing additional history. Overall, the increasing-price trajectory indicates that longer histories help LLM agents coordinate their expectations more tightly and align them more closely with the theoretical benchmark, though convergence to FEE is incomplete. Notably, this challenge is amplified under stronger network effects (bottom row), where dispersion remains higher and the mean shows greater deviation from FEE.

\input{contents/figures/results_gpt_5_increasing}

We next turn to the decreasing-price trajectory (Figure \ref{fig:results_gpt_5_decreasing}), where prices fall steadily from high to low. Here, longer histories largely reduce the dispersion of expectations in this direction, indicating that agents make greater use of historical information to coordinate with each other. However, unlike the increasing-price case, the mean expectations remain consistently below the FEE line, showing systematic underestimation even when the history window is long. This bias persists across network effect strengths, and becomes more pronounced under $\beta=0.75$, where both dispersion and deviation from the benchmark are exacerbated. Taken together, the decreasing-price trajectory highlights that while history improves coordination, convergence to FEE is obstructed by persistent pessimism, and again, stronger network effects make this convergence even more difficult.

\input{contents/figures/results_gpt_5_decreasing}

\subsubsection{Non-Monotonic Price Trajectories}

So far, we have focused on monotonic price changes, where the trajectory provides a simple directional tendency. A natural question is whether more complex price patterns might also influence how LLM agents form expectations. In particular, non-monotonic sequences can obscure the information structure available to the agents: rather than learning from a clear upward or downward trend, agents must interpret less predictable patterns. This design allows us to test whether LLM agents can still extract meaningful information from history and whether such patterns affect their convergence to the FEE benchmark, especially under stronger network effects.

Figure~\ref{fig:results_gpt_5_converging} reports results for the converging sequence, where prices start at an extreme and gradually move toward the mean.  Without loss of generality, we use $\{49.99, \allowbreak 37.49, \allowbreak 47.49, \allowbreak 39.99, \allowbreak 44.99, \allowbreak 42.49\}$ in the experiment. Compared with the monotonic cases, agent performance is weaker. With only a short history window, expectations are highly dispersed, and the average deviates substantially from the FEE line, often performing worse than the static baseline. Extending the window to three and six rounds narrows dispersion somewhat, but the mean remains noticeably off the benchmark even with long histories. This shows that without a clear directional signal, LLM agents struggle to make effective use of history, and stronger network effects amplify this difficulty by widening dispersion and increasing the gap from FEE.

\input{contents/figures/results_gpt_5_converging}

By contrast, Figure~\ref{fig:results_gpt_5_diverging} 
presents the diverging sequence, where prices begin near the mean and gradually move toward extremes. Specifically, we used $\{42.49, \allowbreak 44.99, \allowbreak 39.99, \allowbreak 47.49, \allowbreak 37.49, \allowbreak 49.99\}$ Here too, agent performance is notably weaker than in the monotonic cases. With short histories, expectations are scattered and biased away from the FEE line, and while longer histories reduce dispersion, the improvement is modest. The group’s mean continues to drift from the benchmark at both ends of the price range. As in the converging case, stronger network effects exacerbate the problem, producing even greater instability. Taken together, these results highlight that when faced with non-monotonic price paths, history provides limited coordination benefits, and increasing network interdependence makes convergence to FEE more difficult rather than easier.

\input{contents/figures/results_gpt_5_diverging}

\subsection{Quantitative Analysis on Agent Performance}

To move beyond visual comparisons, we quantify the gap between LLM agents and the FEE using Root Mean Squared Error (RMSE). For each experimental cell—defined by a network-effect strength $\beta$, a price trajectory (static, increasing, decreasing, converging, or diverging), and a history window length—we observe $M$ rounds (six price points in our design) and $N$ agents (here $N=50$). Let $\hat{y}_{i}(p_j)$ denote agent $i$’s stated expectation of total participants at price $p_j$, and let $y_{\text{FEE}}(p_j)$ be the corresponding theoretical benchmark implied by FEE under the same $\beta$ and type distribution. Note that, by construction, $y_{\text{FEE}}(p_j)$ does not depend on $i$ because all economic agents will share the same expectation under each price point (see Section \ref{sec:game} for detail). We define the RMSE for a given cell as

\begin{equation}
\text{RMSE} \;=\; \sqrt{\frac{1}{MN} \sum_{j=1}^{M} \sum_{i=1}^{N} \Big( y_{\text{FEE}}(p_j) - \hat{y}_{i}(p_j) \Big)^2 },
\end{equation}

A smaller RMSE indicates closer alignment with the FEE benchmark.

\input{contents/figures/metrics_rmse_gpt_5}

Figure~\ref{fig:metric_gpt_5_overview} presents LLM agents' RMSE across all scenarios. Panel (A) shows the static benchmark: when agents are given no history, the gap from FEE is already notable, with RMSE = 6.712 under weak network effects ($\beta=0.25$) and doubling to 13.979 under strong network effects ($\beta=0.75$). This confirms that network effects alone could make coordination more fragile.

Panels (B) and (C) report results under monotonic price trajectories. For decreasing prices (Panel B), the error is very high when history is short (RMSE = 9.031 and 17.203 for weak and strong $\beta$). Expanding the history window steadily reduces error, reaching 2.486 (weak) and 5.026 (strong) by thirteen rounds. Increasing prices (Panel C) shows a similar pattern: RMSE falls sharply as history grows from one to seven rounds, with only small additional improvement by thirteen rounds (e.g., from 6.180 to 2.154 under weak $\beta$, and from 11.136 to 5.242 under strong $\beta$). These results confirm that history helps LLM agents coordinate more effectively, though convergence to FEE remains incomplete and more difficult when network effects are stronger.

Panels (D) and (E) show the non-monotonic cases. In converging prices (Panel D), RMSE remains persistently high, especially under strong $\beta$: error starts at 18.731 and only decreases to 13.884 even with long histories. For weak $\beta$, the improvement is greater (from 15.912 to 6.534), but still weaker than in the monotonic cases. In diverging prices (Panel E), performance improves more with longer memory (e.g., from 17.211 to 4.301 under weak $\beta$, and from 18.713 to 9.120 under strong $\beta$), but agents still fail to align fully with FEE, particularly under strong network effects. These findings highlight that when the price trajectory lacks a clear monotonic signal, history provides limited coordination benefits.

Overall, three consistent patterns emerge from Figure~\ref{fig:metric_gpt_5_overview}. First, stronger network effects systematically increase deviations from FEE, regardless of trajectory or memory length. Second, extending history improves performance, but the marginal gain diminishes beyond a moderate window. Third, non-monotonic price sequences sustain substantially higher errors than monotonic ones, revealing that complexity in the environment exacerbates coordination failures. Together, these results demonstrate that while memory helps, network strength and trajectory complexity fundamentally constrain LLM agents’ ability to replicate the equilibrium reasoning assumed in economic theory.

\section{Individual-Level Analysis}\label{sec:regression}

The previous analysis evaluated LLM agents’ collective deviations from the fulfilled expectation equilibrium (FEE) using group-level RMSE. We now move one step deeper and ask: how do individual LLM agents deviate from the FEE benchmark, and what systematic factors drive these deviations? To capture this deviation, we define
\[
Y \;=\; \hat{y}(p) - y_{\text{FEE}}(p),
\]
where $\hat{y}(p)$ is the stated expectation of the number of participants at price $p$, and $y_{\text{FEE}}(p)$ is the theoretical prediction under the same price. By construction, $Y=0$ should always hold for fully rational economic agents.

The estimation sample contains $7{,}800$ observations from the full-factorial experimental design. The regressors include the posted price ($Price$), the strength of network effects ($NE\in\{0,1\}$, indicating weak or strong), the standalone value of the agent ($\theta$), and the accessible history window ($History \in \{0,0.5,1\}$, corresponding to 1, 7, 13 rounds of history used previously, respectively). Because $Y$ can take both positive and negative values and exhibits heavy tails, we apply a Yeo–Johnson transformation to stabilize inference. All regressions include price-path fixed effects (static, increasing, decreasing, converging, and diverging trajectories).

To systematically investigate the drivers of bias, we estimate four nested OLS models. Model~1 includes the main effects of $Price$, $NE$, $\theta$, and $History$, together with fixed effects for 4 price trajectories. Model~2 augments this baseline with an interaction between $Price$ and $\theta$, capturing how sensitivity to price depends on agent type. Model~3 introduces interactions between $NE$ and the other regressors, allowing us to test whether the presence of stronger network effects systematically amplifies or dampens deviations. Finally, Model~4 replaces these with interactions between $History$ and the other regressors, to assess whether longer history reshapes the influence of price and type on expectations. Taken together, these four models provide a comprehensive view of how internal heterogeneity and external conditions shape LLM agents’ deviations from FEE.

\begin{align*}
\text{Model 1:}\quad 
Y &= \beta_0 + \beta_1\times Price + \beta_2 \times NE \notag\\
  & + \beta_3 \times \theta + \beta_4 \times History \notag\\
  & + \boldsymbol{\gamma} \mathbf{FE} + \varepsilon,
\end{align*}

\begin{align*}
  \text{Model 2:}\quad 
Y &= \beta_0 + \beta_1\times Price + \beta_2 \times NE \notag\\
  & + \beta_3 \times \theta + \beta_4 \times History \notag\\
  &+ \beta_5\times (Price\cdot\theta)\\
  & + \boldsymbol{\gamma} \mathbf{FE} + \varepsilon,
\end{align*}

\begin{align*}
    \text{Model 3:}\quad 
Y &= \beta_0 + \beta_1\times Price + \beta_2 \times NE \notag\\
  & + \beta_3 \times \theta + \beta_4 \times History \notag\\
  &+ \beta_5\times (Price\cdot\theta)\\
  &+ \beta_6\times (NE\cdot History)\\
  &+ \beta_7\times (NE\cdot Price)\\
  & + \beta_8\times (NE\cdot \theta)+ \boldsymbol{\gamma} \mathbf{FE} + \varepsilon,
\end{align*}

\begin{align*}
      \text{Model 4:}\quad 
Y &= \beta_0 + \beta_1\times Price + \beta_2 \times NE \notag\\
  & + \beta_3 \times \theta + \beta_4 \times History \notag\\
    &+ \beta_5\times (Price\cdot\theta)\\
  &+ \beta_6\times (NE\cdot History)\\
  &+ \beta_9\times (Price\cdot History)\\
  & + \beta_{10}\times (\theta\cdot History)+ \boldsymbol{\gamma} \mathbf{FE} + \varepsilon.
\end{align*}

\input{contents/individual_analysis_gpt_5}

The regression results are reported in Table~\ref{tab:main4}. The clearest message is that $Price$ is the dominant driver of deviation: across all specifications, the coefficient on $Price$ is large, highly significant, and consistently positive. This implies that higher prices are strongly associated with systematic \emph{overestimation} relative to FEE. The standalone value $\theta$ and $History$ also exert positive and significant main effects (3.671 and 2.056 in Model 1, respectively), suggesting that agents with higher standalone values and longer memory windows expect more participants than FEE predicts. These results corroborate prevailing technical perspectives that LLM-based agent behavior is strongly shaped by configuration—through both external environmental conditions and internal agent settings \citep{wang2023voyager, brown2020language}.

By contrast, the main effect of $NE$ is consistently insignificant (reaching significance only in Model 3), an unexpected but informative finding. This result indicates that network effects do not directly shift expectations in either direction once other covariates are controlled. Instead, as the interaction models show, their influence operates conditionally.

Turning to interaction terms, Model 2 first highlights the strong and negative $Price \times \theta$ effect. This indicates that while higher prices drive overestimation, this tendency is substantially moderated among agents with larger $\theta$ values.

Model 3 then explores how network effects interact with other drivers. The results show that $NE$ significantly amplifies the price-driven bias ($NE \times Price = 14.294$) while simultaneously dampening overestimation among higher-type agents ($NE \times \theta = -1.844$). These findings reveal that network effects are not independent forces of distortion. Instead, they amplify deviations induced by external conditions like price, while conditioning the role of agent heterogeneity. This provides statistical evidence that the interdependency among LLM agents translates into larger systematic deviations only when coupled with contextual pressures.

Finally, Model 4 underscores the dual role of $History$. Its main effect raises expectations, consistent with Model 1, but its interaction terms show significant negative moderation: longer memory reduces the marginal influence of both $Price$ and $\theta$ on deviations. This result reinforces a key message from the earlier RMSE analysis: providing richer historical context narrows dispersion and lowers error, even if full convergence to FEE is not achieved.

Taken together, the individual-level regressions connect directly to the group-level RMSE patterns in Figure~\ref{fig:metric_gpt_5_overview}. The increase in RMSE under strong network effects can now be traced to conditional amplification—particularly through the $NE \times Price$ channel—rather than to a direct, unconditional impact of $NE$. Likewise, the stabilizing influence of history at the aggregate level reflects its moderating role in these regressions, where it tempers sensitivity to price and heterogeneity as memory grows. This finding extends recent work on historical context in AI-agent decision-making \citep{guo2024economics, huang2024far} by identifying and quantifying the moderating role of decision history in shaping agent rationality. In combination, the individual- and group-level results provide a coherent account of why and when LLM agents diverge from FEE.

\section{Conclusion}\label{sec:conclusion}

The growing deployment of AI agents in markets and organizations raises a fundamental question: how do machine agents behave when embedded in interdependent environments? This study provides one of the first systematic answers by examining LLM-based agents in a canonical network-effect game. Using a comprehensive experimental design, we documented how agents respond to network effects, how they use history to form expectations, and how these responses generate predictable deviations from FEE. Our results consistently show that while LLM agents exhibit partial convergence to equilibrium under simple conditions, their behavior systematically departs from classical economic predictions once history and interdependence are introduced.

Network effects lie at the heart of this phenomenon. In economics, network effects create interdependencies that make equilibrium selection both fragile and path-dependent. In our experiments, we found that stronger network effects do not directly shift expectations, but instead magnify deviations induced by external conditions such as price and agent heterogeneity. This amplification produces systematic bias: agents become more optimistic at high prices and more pessimistic at low prices. Such findings demonstrate that network effects remain the central structural force shaping behavior in machine-agent systems, just as they are in human-agent systems, but that their influence operates through conditional amplification rather than unconditional shifts.

The theoretical implications of these findings are profound. Rational-expectations theory rests on the principle that history is irrelevant: past outcomes are sunk and equilibrium is defined by forward-looking consistency. Our evidence shows that LLM agents violate this principle not incidentally but structurally. Because their architecture builds predictions from historical tokens, history is the raw material of their reasoning, not background noise. This explains why agents align partially with equilibrium under simple monotonic trajectories—where history encodes a clear directional signal—but fail systematically under non-monotonic trajectories that scramble the predictive structure. In other words, what is “irrelevant” in classical equilibrium becomes decisive for LLM agents. This evidence pushes theory forward by showing that equilibrium reasoning must be reinterpreted when applied to machine agents: expectation formation is no longer a matter of solving fixed-point equations on beliefs, but of understanding how architectures transform past information into forecasts. This shift reframes equilibrium analysis in a way that bridges economics and AI, and it opens the door to a history-aware game theory of machine collectives.

The practical implications are equally critical. As organizations increasingly rely on autonomous AI agents to make or support decisions in finance, platforms, and operations, the configuration of multi-agent systems cannot be treated as neutral. Our study shows that seemingly minor design choices—such as how much history is accessible, or how interdependencies are encoded—systematically shape collective outcomes. Providing agents with richer historical information narrows dispersion and stabilizes coordination, while leaving them with limited memory creates volatility and bias. Managers and system designers must therefore treat memory and interdependence not as background parameters but as levers that determine whether collective behavior aligns with or diverges from intended outcomes.

The richness of the network-effect context also points toward a much broader agenda. Network effects are only one form of interdependence; others include congestion, complementarities, and reputation spillovers. By showing that LLM agents systematically diverge from equilibrium predictions even in a stylized network-effect game, we establish a blueprint for extending analysis to these other forms of interdependence. Doing so will open up a new research frontier: understanding how machine agents collectively reason, mis-reason, and coordinate in environments where individual actions feed back on others in complex ways.

Another important lesson concerns the heterogeneity of LLM behavior. Our robustness checks across GPT-5 and Qwen3-Plus highlight that while broad patterns persist, the magnitude and direction of deviations can vary across models and generations. This suggests that “machine rationality” is not a stable property, but depends on architectural design choices and training data. Our success in isolating history as a decisive factor is precisely because history plays a central role in transformer architectures, even though it is theoretically sunk for economists. This sharp contrast between machine and human rationality underscores the need for deeper theorizing at the intersection of economics and AI.

Finally, our work comes with limitations that suggest multiple avenues for extension. We focused on stylized network-effect games and controlled communication protocols, while real-world environments involve richer payoff structures, more complex histories, and heterogeneous forms of interdependence. Future research can extend our framework to richer market games, explore interventions such as adaptive prompts or regulation of memory, and test whether different foundation model architectures exhibit qualitatively different strategic behaviors. By doing so, scholars can build on our foundation to develop a general theory of AI-agent interaction in interdependent systems.

\bibliographystyle{apalike}
\bibliography{literature.bib}

\end{document}

%% file: contents/figures/finite_state_machine.tex
\begin{figure}[htbp]
    \centering
    \includegraphics[width=\textwidth]{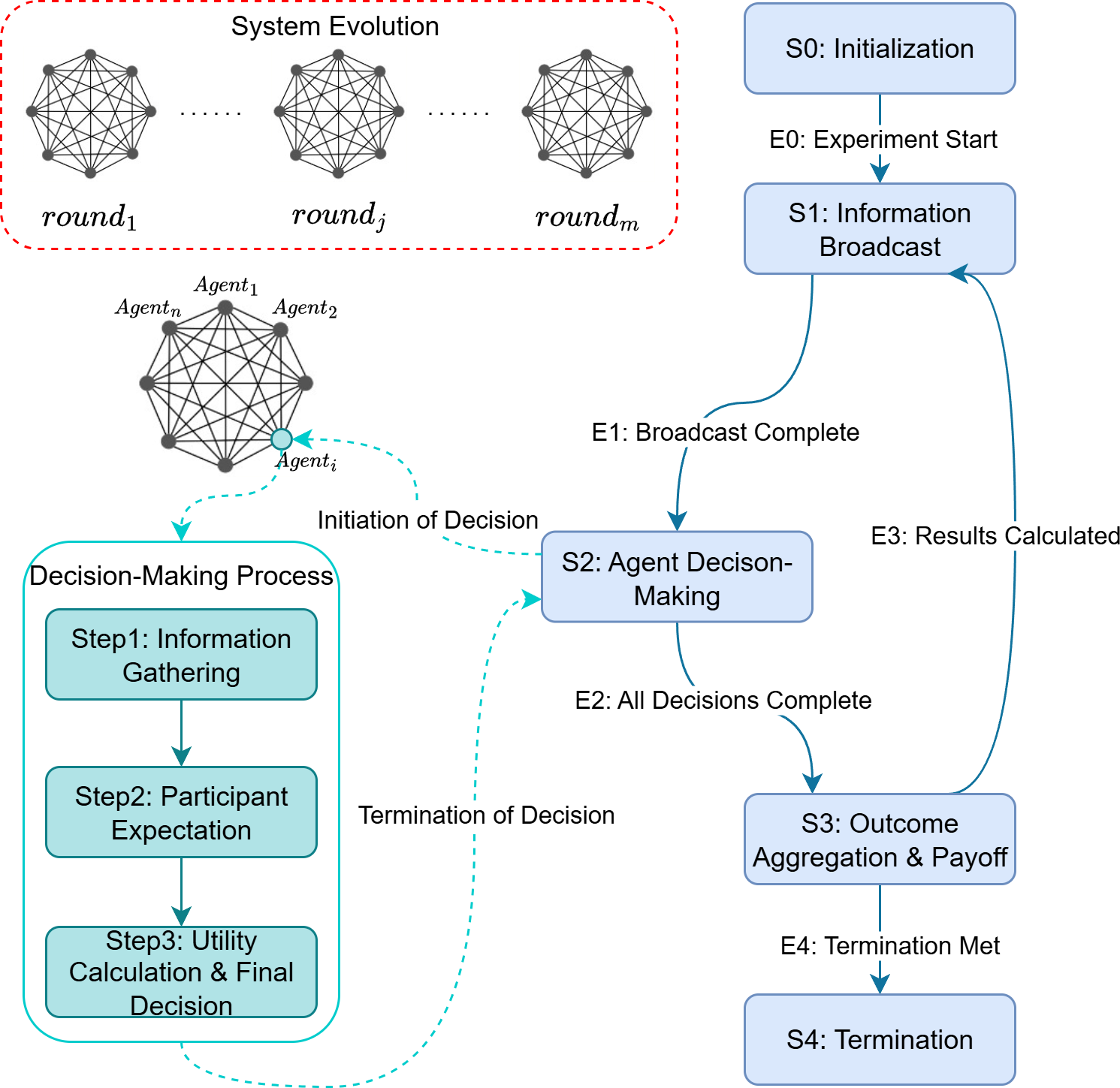}
    \caption{Experimental Workflow.}
    \label{fig:fsm}
\end{figure}

%% file: contents/figures/results_benchmark.tex
\begin{figure}[htbp]
    \centering
    \includegraphics[width=\textwidth]{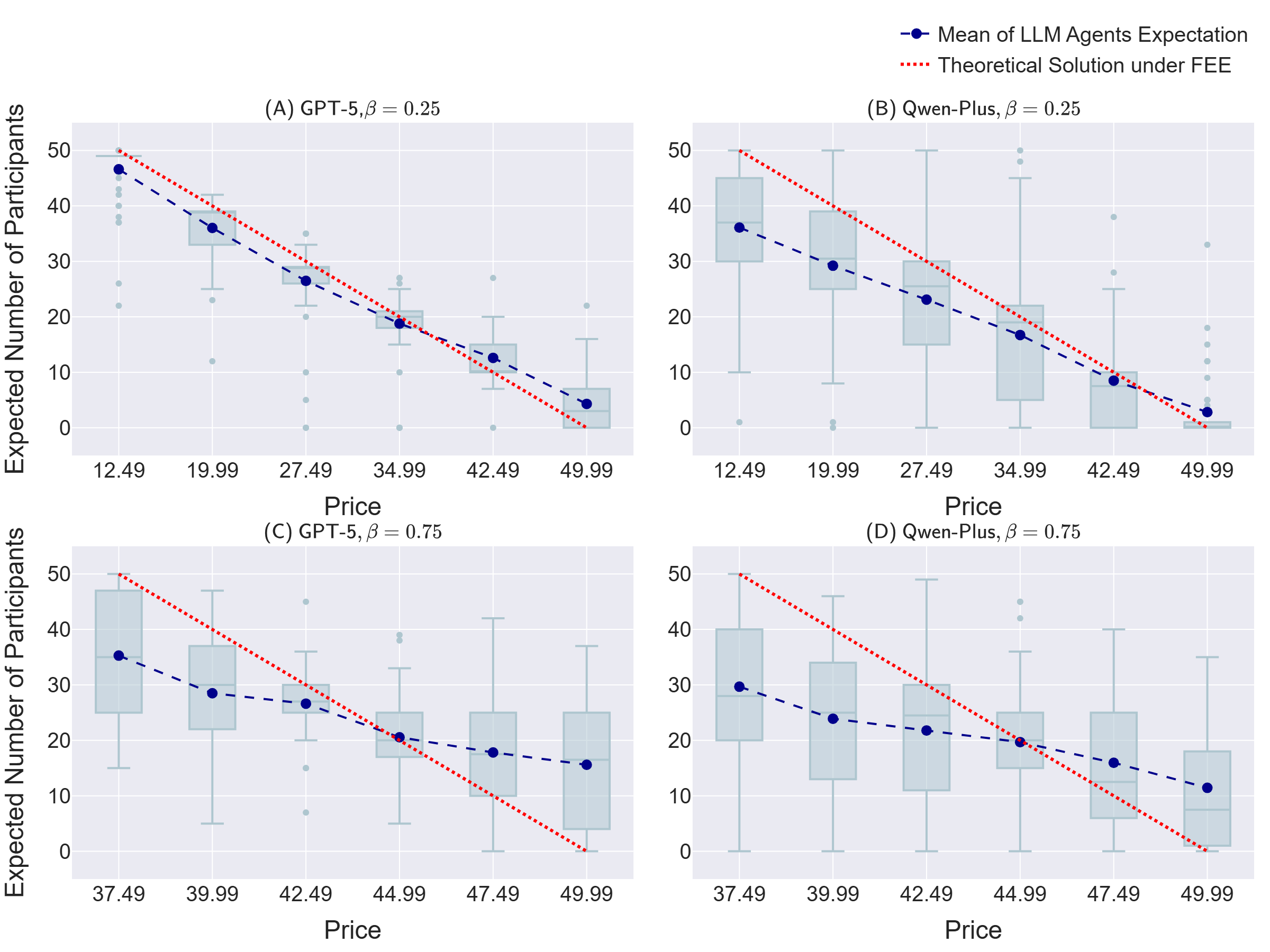}
    \caption{Benchmark: Static Scenario.}
    \label{fig:results_benchmark}
\end{figure}

%% file: contents/figures/results_gpt_5_increasing.tex
\begin{figure}[htbp]
    \centering
    \includegraphics[width=\textwidth]{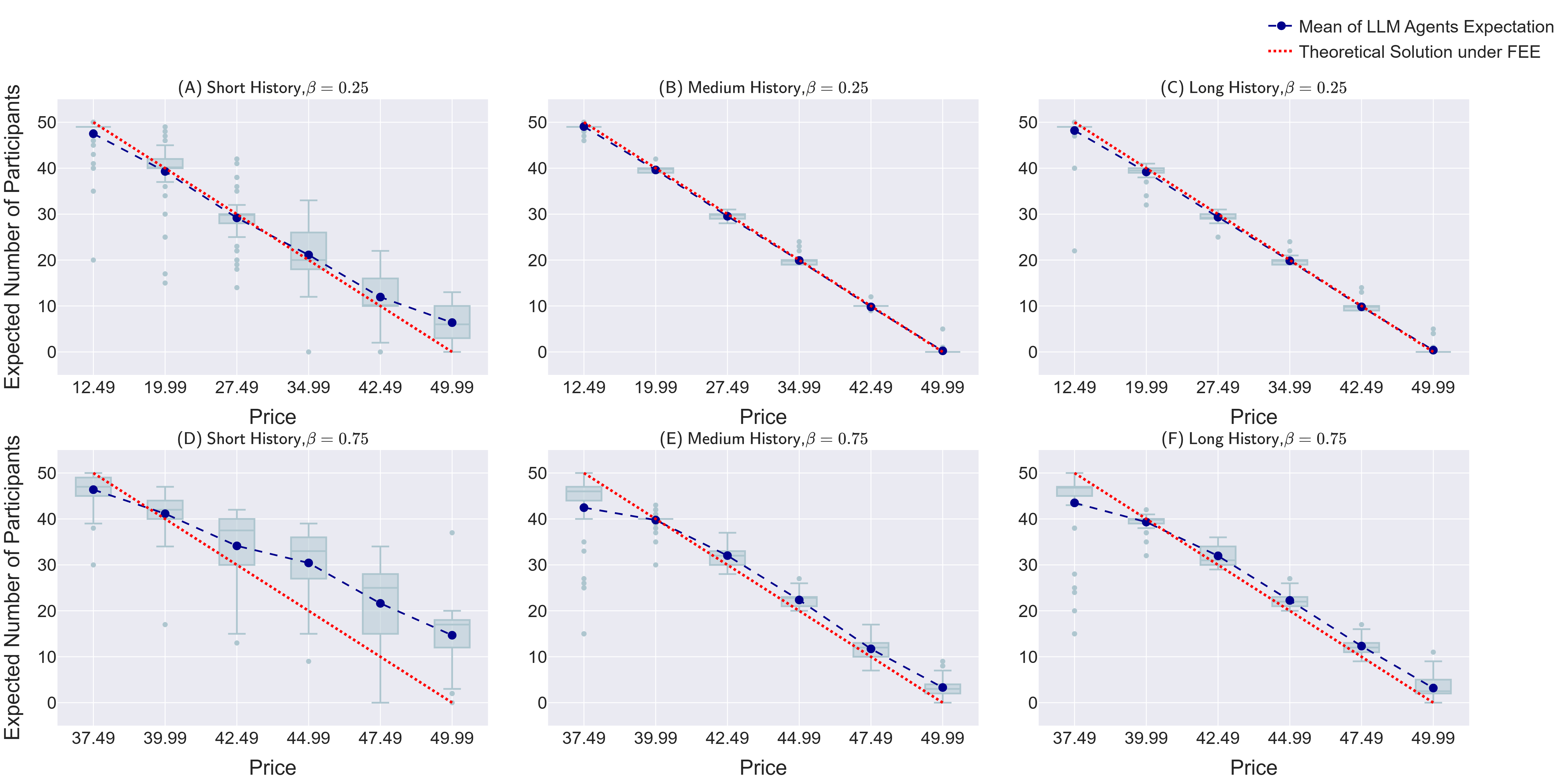}
    \caption{GPT-5: Increasing Prices.}
    \label{fig:results_gpt_5_increasing}
\end{figure}

%% file: contents/figures/results_gpt_5_decreasing.tex
\begin{figure}[htbp]
    \centering
    \includegraphics[width=\textwidth]{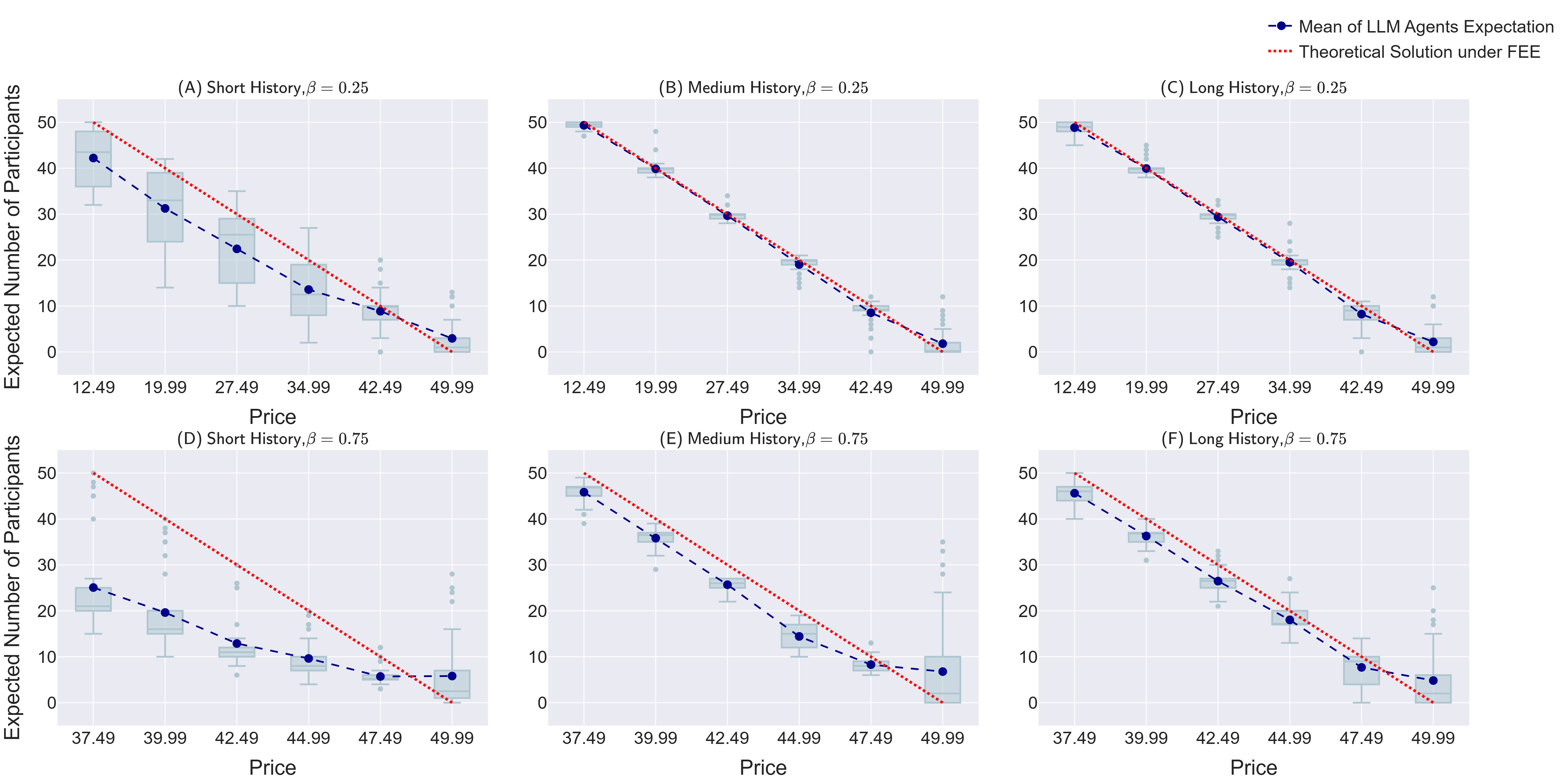}
    \caption{GPT-5: Decreasing Prices.}
    \label{fig:results_gpt_5_decreasing}
\end{figure}

%% file: contents/figures/results_gpt_5_converging.tex
\begin{figure}[htbp]
    \centering
    \includegraphics[width=\textwidth]{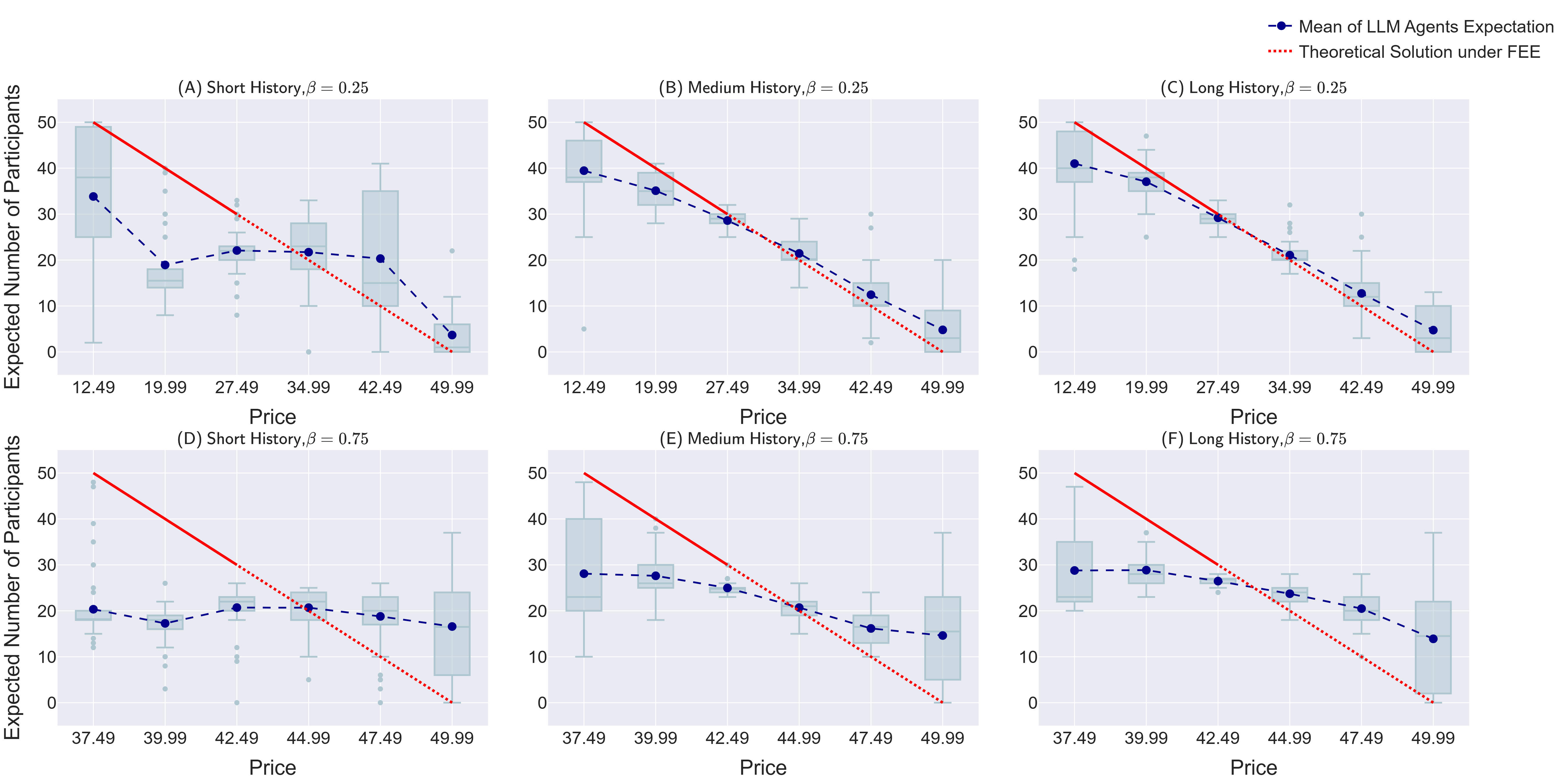}
    \caption{GPT-5: Converging Prices.}
    \label{fig:results_gpt_5_converging}
\end{figure}

%% file: contents/figures/results_gpt_5_diverging.tex
\begin{figure}[htbp]
    \centering
    \includegraphics[width=\textwidth]{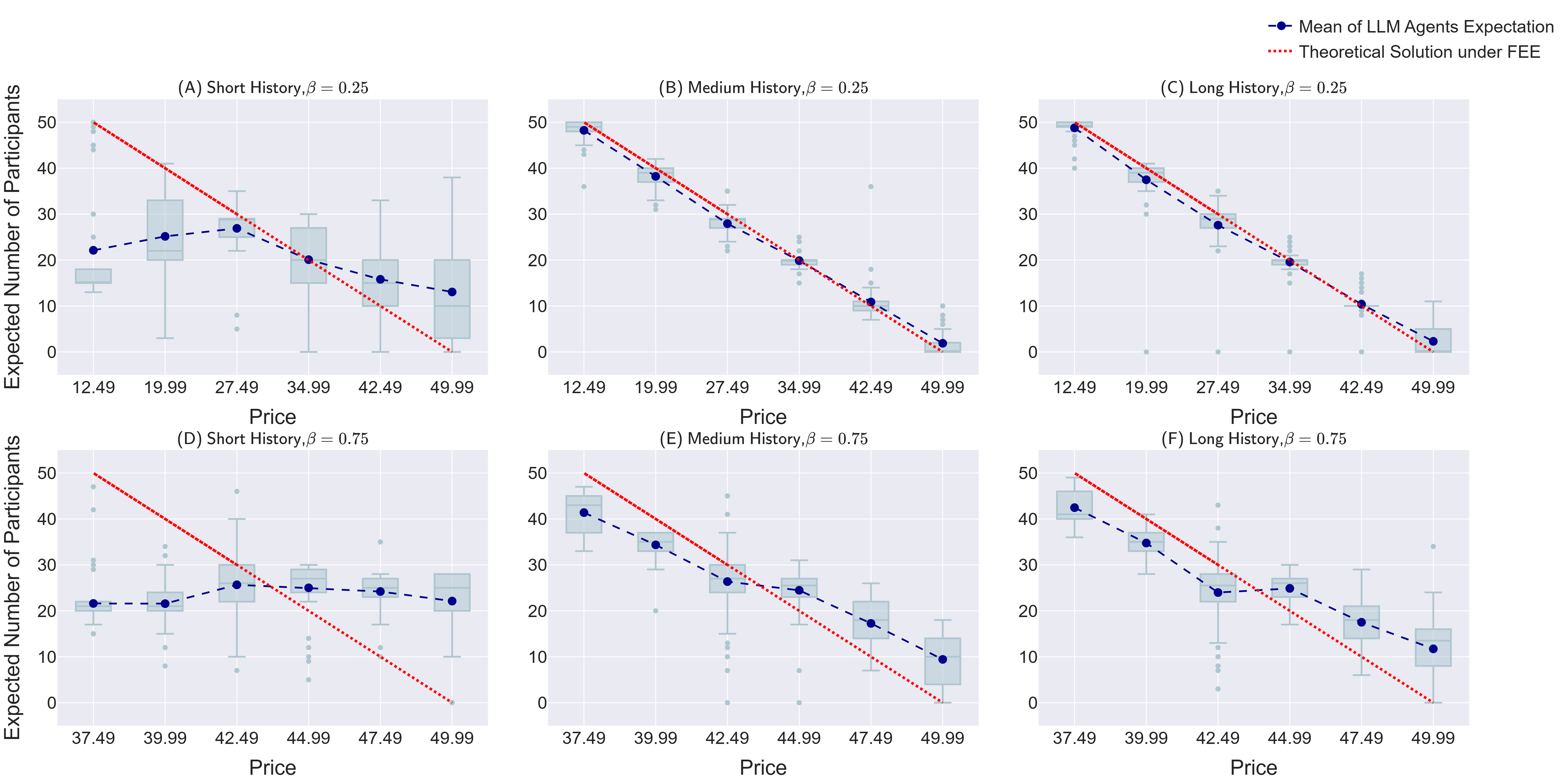}
    \caption{GPT-5: Diverging Prices.}
    \label{fig:results_gpt_5_diverging}
\end{figure}

%% file: contents/figures/metrics_rmse_gpt_5.tex
\begin{figure}[htbp]
    \centering
    \includegraphics[width=\textwidth]{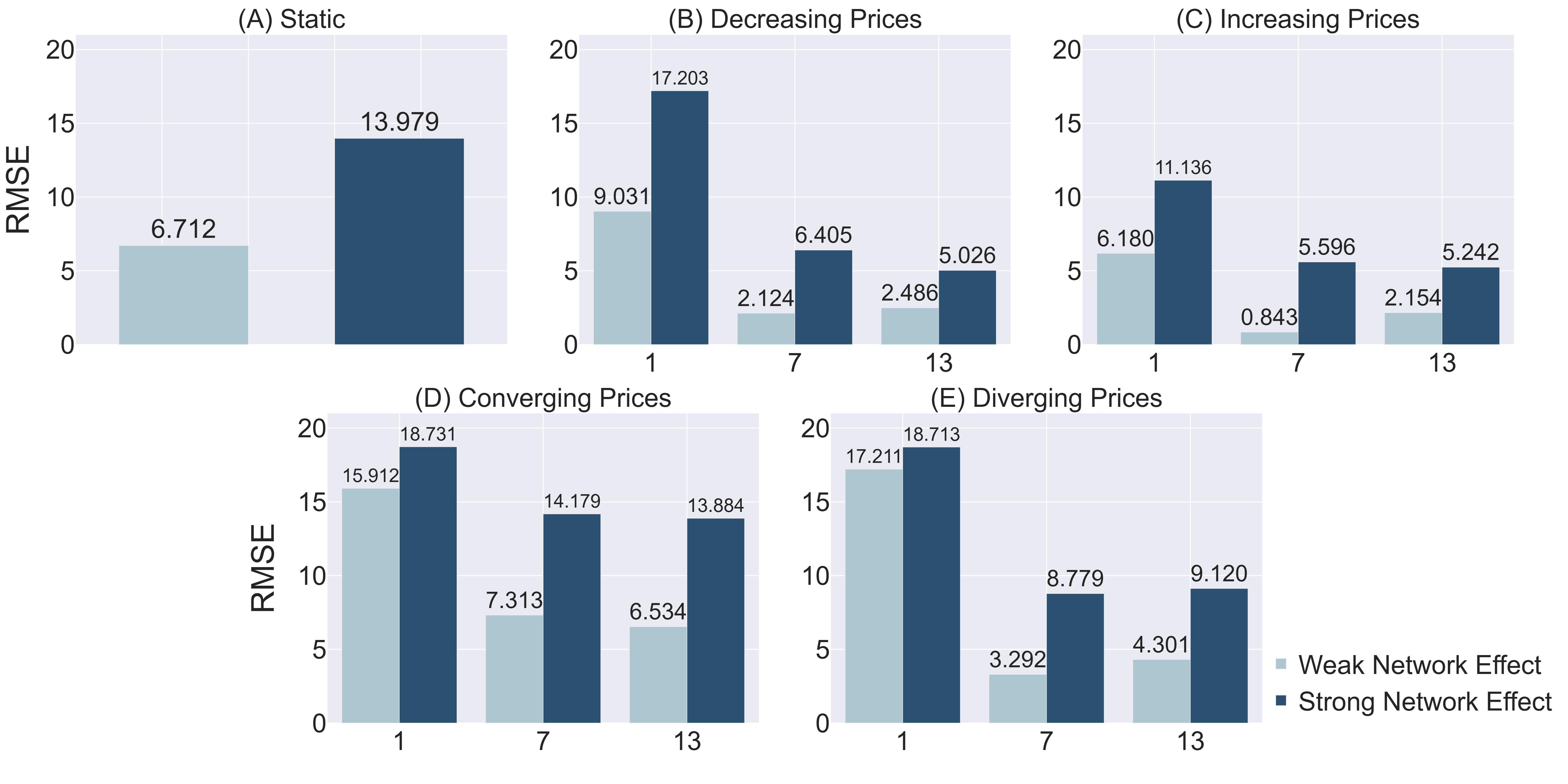}
    \caption{Metric: GPT-5.}
    \label{fig:metric_gpt_5_overview}
\end{figure}

%% file: contents/individual_analysis_gpt_5.tex
\renewcommand\arraystretch{1.0}   %
\setlength{\tabcolsep}{6pt}       %
\renewcommand\cellalign{cc}       %
\newcommand{\coefse}[2]{\makecell{#1\\(#2)}}
\newcommand{\dash}{\makecell{---}}   %

\begin{table}[htbp]
\centering
\caption{Regression Analysis with Price-Path Fixed Effects}
\label{tab:main4}
\begin{tabular}{lcccc}
\toprule
Variable & Model 1 & Model 2 & Model 3 & Model 4 \\
\midrule
$Price$
& \coefse{17.310***}{0.313} 
& \coefse{20.105***}{0.620} 
& \coefse{12.957***}{0.620} 
& \coefse{27.376***}{0.713} \\
$NE$
& \coefse{0.324}{0.180}   
& \coefse{0.324}{0.180}   
& \coefse{-5.836***}{0.552}   
& \coefse{0.408}{0.307}   \\
$\theta$  
& \coefse{3.671***}{0.311}  
& \coefse{6.523***}{0.617}  
& \coefse{7.445***}{0.668}  
& \coefse{9.379***}{0.734}  \\
$History$               
& \coefse{2.056***}{0.247}  
& \coefse{2.056***}{0.247}  
& \coefse{2.148***}{0.299}  
& \coefse{13.057***}{0.620}  \\
$Price \times \theta$ 
& \dash                     
& \coefse{-5.704***}{1.065} 
& \coefse{-5.704***}{1.031} 
& \coefse{-5.704***}{1.009} \\
$NE$ $\times$ $Price$     
& \dash                     
& \dash                     
& \coefse{14.294***}{0.586} 
& \dash                     \\
$NE \times \theta$    
& \dash                     
& \dash                     
& \coefse{-1.844***}{0.604}
& \dash                     \\
$NE$ $\times$ $History$   
& \dash
& \dash
& \coefse{-0.182}{0.430}
& \coefse{-0.182}{0.428} \\
$Price$ $\times$ $History$ 
& \dash
& \dash
& \dash
& \coefse{-15.755***}{0.731}\\
$\theta$ $\times$ $History$ 
& \dash
& \dash
& \dash
& \coefse{-6.189***}{0.734} \\
\midrule
\textit{Price-path fixed effects} & \multicolumn{4}{c}{Included} \\
\# Obs  & 7,800 & 7,800 & 7,800 & 7,800 \\
$R^2$   & 0.385 & 0.388 & 0.447 & 0.442 \\
\bottomrule
\multicolumn{5}{l}{\footnotesize Notes: *** $p<0.01$; ** $p<0.05$. HC3 robust SEs in parentheses.}
\end{tabular}
\end{table}

%% file: Manuscript.bbl
\begin{thebibliography}{}

\bibitem[Abbasi et~al., 2024]{abbasi2024pathways}
Abbasi, A., Parsons, J., Pant, G., Sheng, O. R.~L., and Sarker, S. (2024).
\newblock Pathways for design research on artificial intelligence.
\newblock {\em MIS Quarterly}.
\newblock Forthcoming.

\bibitem[Achiam et~al., 2023]{achiam2023gpt}
Achiam, J., Adler, S., Agarwal, S., Ahmad, L., Akkaya, I., Aleman, F.~L.,
  Almeida, D., Altenschmidt, J., Altman, S., Anadkat, S., et~al. (2023).
\newblock Gpt-4 technical report.
\newblock {\em arXiv preprint arXiv:2303.08774}.

\bibitem[Akata et~al., 2025]{akata2025playing}
Akata, E., Schulz, L., Coda-Forno, J., Oh, S.~J., Bethge, M., and Schulz, E.
  (2025).
\newblock Playing repeated games with large language models.
\newblock {\em Nature Human Behaviour}, pages 1--11.

\bibitem[Anthis et~al., 2025]{anthis2025llm}
Anthis, J.~R., Liu, R., Richardson, S.~M., Kozlowski, A.~C., Koch, B., Evans,
  J., Brynjolfsson, E., and Bernstein, M. (2025).
\newblock Llm social simulations are a promising research method.
\newblock {\em arXiv preprint arXiv:2504.02234}.

\bibitem[Arthur, 1989]{arthur1989competing}
Arthur, W.~B. (1989).
\newblock Competing technologies, increasing returns, and lock-in by historical
  events.
\newblock {\em The economic journal}, 99(394):116--131.

\bibitem[Bikhchandani et~al., 1992]{bikhchandani1992theory}
Bikhchandani, S., Hirshleifer, D., and Welch, I. (1992).
\newblock A theory of fads, fashion, custom, and cultural change as
  informational cascades.
\newblock {\em Journal of political Economy}, 100(5):992--1026.

\bibitem[Boudreau, 2021]{boudreau2021promoting}
Boudreau, K.~J. (2021).
\newblock Promoting platform takeoff and self-fulfilling expectations: Field
  experimental evidence.
\newblock {\em Management Science}, 67(9):5953--5967.

\bibitem[Brown et~al., 2020]{brown2020language}
Brown, T., Mann, B., Ryder, N., Subbiah, M., Kaplan, J.~D., Dhariwal, P.,
  Neelakantan, A., Shyam, P., Sastry, G., Askell, A., et~al. (2020).
\newblock Language models are few-shot learners.
\newblock {\em Advances in neural information processing systems},
  33:1877--1901.

\bibitem[Camerer et~al., 2004]{camerer2004cognitive}
Camerer, C.~F., Ho, T.-H., and Chong, J.-K. (2004).
\newblock A cognitive hierarchy model of games.
\newblock {\em The quarterly journal of economics}, 119(3):861--898.

\bibitem[Chen et~al., 2023]{chen2023emergence}
Chen, Y., Liu, T.~X., Shan, Y., and Zhong, S. (2023).
\newblock The emergence of economic rationality of gpt.
\newblock {\em Proceedings of the National Academy of Sciences},
  120(51):e2316205120.

\bibitem[Conner, 1995]{conner1995imitated}
Conner, K.~R. (1995).
\newblock Obtaining strategic advantage from being imitated: When can clones
  succeed?
\newblock {\em Management Science}, 41(2):209--225.

\bibitem[Deng et~al., 2025]{deng2025can}
Deng, Z. et~al. (2025).
\newblock Can llm agents recognize demographic heterogeneity in economic games?
\newblock {\em SSRN Electronic Journal}.
\newblock Available at SSRN: \url{https://ssrn.com/abstract=5197303}.

\bibitem[Dou et~al., 2025]{dou2025ai}
Dou, W.~W., Goldstein, I., and Ji, Y. (2025).
\newblock Ai-powered trading, algorithmic collusion, and price efficiency.
\newblock {\em Jacobs Levy Equity Management Center for Quantitative Financial
  Research Paper, The Wharton School Research Paper}.

\bibitem[Economides, 1996]{economides1996economics}
Economides, N. (1996).
\newblock The economics of networks.
\newblock {\em International Journal of Industrial Organization},
  14(6):673--699.

\bibitem[Fan et~al., 2024]{fan2024can}
Fan, C., Chen, J., Jin, Y., and He, H. (2024).
\newblock Can large language models serve as rational players in game theory? a
  systematic analysis.
\newblock In {\em Proceedings of the AAAI Conference on Artificial
  Intelligence}, volume~38, pages 17960--17967.

\bibitem[Farrell and Saloner, 1985]{farrell1985standardization}
Farrell, J. and Saloner, G. (1985).
\newblock Standardization, compatibility, and innovation.
\newblock {\em The Quarterly Journal of Economics}, 100(1):70--83.

\bibitem[Fontana et~al., 2025]{fontana2025nicer}
Fontana, N., Pierri, F., and Aiello, L.~M. (2025).
\newblock Nicer than humans: How do large language models behave in the
  prisoner's dilemma?
\newblock In {\em Proceedings of the International AAAI Conference on Web and
  Social Media}, volume~19, pages 202--213.

\bibitem[Gandhi et~al., 2023]{gandhi2023strategic}
Gandhi, K., Sadigh, D., and Goodman, N.~D. (2023).
\newblock Strategic reasoning with language models.
\newblock {\em arXiv preprint arXiv:2305.19165}.

\bibitem[Gnewuch et~al., 2024]{gnewuch2024more}
Gnewuch, U., Morana, S., Hinz, O., Kellner, R., and Maedche, A. (2024).
\newblock More than a bot? the impact of disclosing human involvement on
  customer interactions with hybrid service agents.
\newblock {\em Information Systems Research}.
\newblock Published Online.

\bibitem[Guo et~al., 2024]{guo2024economics}
Guo, S., Bu, H., Wang, H., Ren, Y., Sui, D., Shang, Y., and Lu, S. (2024).
\newblock Economics arena for large language models.
\newblock {\em arXiv preprint arXiv:2401.01735}.

\bibitem[Han et~al., 2023]{han2023bots}
Han, E., Yin, D., and Zhang, H. (2023).
\newblock Bots with feelings: Should ai agents express positive emotion in
  customer service?
\newblock {\em Information Systems Research}.
\newblock Published Online.

\bibitem[Horton, 2023]{horton2023large}
Horton, J.~J. (2023).
\newblock Large language models as simulated economic agents: What can we learn
  from homo silicus?
\newblock Technical report, National Bureau of Economic Research.

\bibitem[Huang et~al., 2024]{huang2024far}
Huang, J.-t., Wang, Z., Zhang, M., et~al. (2024).
\newblock How far are we on the decision-making of llms? evaluating llms'
  gaming ability in multi-agent environments.
\newblock {\em arXiv preprint arXiv:2403.11807}.

\bibitem[Jing, 2007]{jing2007network}
Jing, B. (2007).
\newblock Network externalities and market segmentation in a monopoly.
\newblock {\em Economics Letters}, 95(1):7--13.

\bibitem[Karten et~al., 2025]{karten2025llm}
Karten, S. et~al. (2025).
\newblock Llm economist: Large population models and mechanism design in
  multi-agent generative simulacra.
\newblock {\em arXiv preprint arXiv:2507.15815}.

\bibitem[Katz and Shapiro, 1985]{katz1985network}
Katz, M.~L. and Shapiro, C. (1985).
\newblock Network externalities, competition, and compatibility.
\newblock {\em The American economic review}, 75(3):424--440.

\bibitem[Katz and Shapiro, 1986]{katz1986technology}
Katz, M.~L. and Shapiro, C. (1986).
\newblock Technology adoption in the presence of network externalities.
\newblock {\em The Journal of Political Economy}, 94(4):822--841.

\bibitem[Lor{\`e} and Heydari, 2024]{lore2024strategic}
Lor{\`e}, N. and Heydari, B. (2024).
\newblock Strategic behavior of large language models and the role of game
  structure versus contextual framing.
\newblock {\em Scientific Reports}, 14(1):18490.

\bibitem[Rahwan et~al., 2019]{rahwan2019machine}
Rahwan, I., Cebrian, M., Obradovich, N., Bongard, J., Bonnefon, J.-F.,
  Breazeal, C., Crandall, J.~W., Christakis, N.~A., Couzin, I.~D., Jackson,
  M.~O., et~al. (2019).
\newblock Machine behaviour.
\newblock {\em Nature}, 568(7753):477--486.

\bibitem[Rosales and Paulitsch, 2021]{rosales2021composable}
Rosales, R. and Paulitsch, M. (2021).
\newblock Composable finite state machine-based modeling for
  quality-of-information-aware cyber-physical systems.
\newblock {\em ACM Transactions on Cyber-Physical Systems}, 5(2):1--27.

\bibitem[Seymour et~al., 2025]{seymour2025less}
Seymour, M., Yuan, L.~I., Riemer, K., and Dennis, A.~R. (2025).
\newblock Less artificial, more intelligent: Understanding affinity,
  trustworthiness, and preference for digital humans.
\newblock {\em Information Systems Research}.
\newblock Published Online.

\bibitem[Silva, 2024]{silva2024large}
Silva, A. (2024).
\newblock Large language models playing mixed strategy nash equilibrium games.
\newblock In {\em International Conference on Network Games, Artificial
  Intelligence, Control and Optimization}, pages 15--27. Springer.

\bibitem[Stodden et~al., 2016]{stodden2016enhancing}
Stodden, V., McNutt, M., Bailey, D.~H., Deelman, E., Gil, Y., Hanson, B.,
  Heroux, M.~A., Ioannidis, J.~P., and Taufer, M. (2016).
\newblock Enhancing reproducibility for computational methods.
\newblock {\em Science}, 354(6317):1240--1241.

\bibitem[Sun et~al., 2025]{sun2025game}
Sun, H., Tang, J., Fan, C., et~al. (2025).
\newblock Game theory meets large language models: A systematic survey.
\newblock {\em arXiv preprint arXiv:2502.09053}.

\bibitem[Taillandier et~al., 2025]{taillandier2025integrating}
Taillandier, P. et~al. (2025).
\newblock Integrating llm in agent-based social simulation: Opportunities and
  challenges.
\newblock {\em arXiv preprint arXiv:2507.19364}.

\bibitem[Wang et~al., 2023]{wang2023voyager}
Wang, G., Xie, Y., Jiang, Y., Mandlekar, A., Xiao, C., Zhu, Y., Fan, L., and
  Anandkumar, A. (2023).
\newblock Voyager: An open-ended embodied agent with large language models.
\newblock {\em arXiv preprint arXiv:2305.16291}.

\bibitem[Willis et~al., 2025]{willis2025will}
Willis, R., Du, Y., Leibo, J.~Z., and Luck, M. (2025).
\newblock Will systems of llm agents cooperate: An investigation into a social
  dilemma.
\newblock {\em arXiv preprint arXiv:2501.16173}.

\bibitem[Xie et~al., 2024]{xie2024can}
Xie, C., Chen, C., Jia, F., Ye, Z., Lai, S., Shu, K., Gu, J., Bibi, A., Hu, Z.,
  Jurgens, D., et~al. (2024).
\newblock Can large language model agents simulate human trust behavior?
\newblock {\em Advances in neural information processing systems},
  37:15674--15729.

\bibitem[Xu et~al., 2024]{xu2024ai}
Xu, R., Sun, Y., Ren, M., Guo, S., Pan, R., Lin, H., Sun, L., and Han, X.
  (2024).
\newblock Ai for social science and social science of ai: A survey.
\newblock {\em Information Processing \& Management}, 61(3):103665.

\bibitem[Yoo et~al., 2024]{yoo2024next}
Yoo, Y., Henfridsson, O., Kallinikos, J., Gregory, R., Burtch, G., Chatterjee,
  S., and Sarker, S. (2024).
\newblock The next frontiers of digital innovation research.
\newblock {\em Information Systems Research}.
\newblock Published Online.

\bibitem[Zuzak et~al., 2011]{zuzak2011finite}
Zuzak, I., Budiselic, I., and Delac, G. (2011).
\newblock A finite-state machine approach for modeling and analyzing restful
  systems.
\newblock {\em Journal of Web Engineering}, pages 353--390.

\end{thebibliography}
